\pdfoutput=1
\documentclass[11pt,letterpaper]{article}

\usepackage[margin=1in,includefoot,bottom=1in]{geometry}

\usepackage{amsmath}
\usepackage{newtxtext,newtxmath}

\usepackage{amssymb}
\usepackage[numbers]{natbib}
\usepackage{booktabs}
\usepackage{array}
\usepackage{multicol}
\usepackage{tikz}
\usepackage{enumitem}
\usepackage{graphicx}
\usepackage[framemethod=default]{mdframed}
\usepackage[colorlinks=true,allcolors=black]{hyperref}

\usepackage{amsthm}
\newtheorem{definition}{Definition}

\usepackage{dsfont}
\newcommand{\ind}[1]{\mathds{1}\{#1\}}
\usepackage[compact]{titlesec}
\titlespacing*{\section}{0pt}{1.6ex plus .2ex minus .2ex}{0.9ex plus .1ex}
\titlespacing*{\subsection}{0pt}{1.2ex plus .2ex minus .2ex}{0.6ex plus .1ex}

\title{\bfseries From One Query to Many:\\A Critical Survey of Privacy Accounting in RAG}
\author{Zhiyuan Zhang\\
\small\texttt{hollyzhang03@ucla.edu}}
\date{}
\AtBeginDocument{%
  \setlength{\abovedisplayskip}{4pt plus 1pt minus 1pt}%
  \setlength{\belowdisplayskip}{4pt plus 1pt minus 1pt}%
  \setlength{\abovedisplayshortskip}{0pt plus 1pt}%
  \setlength{\belowdisplayshortskip}{2pt plus 1pt minus 1pt}%
}
\usepackage{xcolor}

\newcommand{\eps}{\varepsilon}
\newcommand{\nb}{\mathrel{\sim}}

\newcommand{\Dq}[1]{D_q(#1)}
\newcommand{\LLM}[1]{\mathrm{LLM}\!\left(#1\right)}
\newcommand{\topk}{\operatorname{top\text{-}k}}
\begin{document}
\maketitle
\vspace{-2em}

\begin{abstract}
\noindent
Retrieval-augmented generation (RAG) gives a language model access to an external corpus at
inference time, which creates privacy risk when the corpus is sensitive. Differential privacy (DP)
offers formal protection, but the guarantees reported for DP-RAG methods are hard to compare. A
budget certified for one response does not bound a stream of queries, and the same value of $\eps$
can mean different things depending on the protected unit, the accounting horizon, and which
outputs are observable. We divide existing systems into query-time privatization, multi-query
accounting, and one-time corpus privatization; record what each source paper states and evaluates;
and interpret each guarantee along three dimensions: privacy granularity, accounting scope, and
release boundary. A reported $\eps$ is meaningful only when all three are specified.

\end{abstract}

\section{Introduction}
\label{sec:intro}

Retrieval-augmented generation (RAG) lets a language model use an external corpus at inference
time. This is useful in domains whose data were not seen during pretraining, but risky when those
data are sensitive. Without protection, RAG systems leak their corpora: an attacker can infer
whether a document is in the retrieval database
\citep{anderson2024ismydata,liu2024maskmia,naseh2025riddle,li2025generating} and can reconstruct
document content through adversarial prompting or poisoning
\citep{jiang2024feedback,peng_data_2024,chaudhari_phantom_2024}. Several
differentially private RAG (DP-RAG) systems have been proposed in response, each reporting a
privacy budget.

These budgets are not directly comparable. A guarantee for one answer says little about a stream
of answers, and the difference depends on the number of queries, which a single-response guarantee
does not mention. Two systems can also report the same $\varepsilon$ while protecting different
units, accounting over different objects, and exposing different outputs, so equal values of
$\varepsilon$ do not imply equal guarantees.

This survey is organised around the move from one query to many. We group systems by when they
spend the privacy budget and record what each source paper states and evaluates
(Sections~\ref{sec:background}--\ref{sec:taxonomy}, Table~\ref{tab:families}). We then write a
DP-RAG guarantee as one inequality with three free slots (Eq.~\ref{eq:triple}), namely the
neighbouring relation, the certified parameters as a function of the number of queries, and the
observable, and discuss each slot in Section~\ref{sec:dimensions}. Section~\ref{sec:practice}
covers what still prevents comparison when all three agree: query overlap, budget exhaustion, and
the match between attack evaluations and guarantees.

\section{Background}
\label{sec:background}

Let $D$ be a dataset and $\nb$ a symmetric neighbouring relation on datasets. The relation
determines the protected unit, for example one document or one person.

\begin{definition}[Differential privacy \citep{dwork_calibrating_2006,dwork2014algorithmic}]
\label{def:dp}
A randomised mechanism $\mathcal M$ is $(\eps,\delta)$-differentially private with respect to
$\nb$ if, for every pair $D\nb D'$ and every measurable set $S$ of outputs,
\begin{equation}
\label{eq:dp}
\Pr[\mathcal M(D)\in S] \;\le\; e^{\eps}\Pr[\mathcal M(D')\in S]+\delta.
\end{equation}
\end{definition}

Equivalently, DP limits how well any test can distinguish neighbouring inputs. If a test that
observes one output has error rates $\alpha:=\Pr[\text{decide }D'\mid D]$ and
$\beta:=\Pr[\text{decide }D\mid D']$, then for every $D\nb D'$ and every such test
\citep{kairouz_composition_2017-1}
\begin{equation}
\label{eq:ht}
\alpha+e^{\eps}\beta\;\ge\;1-\delta
\qquad\text{and}\qquad
 e^{\eps}\alpha+\beta\;\ge\;1-\delta.
\end{equation}
The mechanism used most often by the systems below is the exponential mechanism
\citep{mcsherry2007mechanism}. Given a utility $u(D,v)$ over a finite candidate set that changes by
at most $\Delta_u$ between neighbours, sampling $v$ with probability proportional to
$\exp\bigl(\eps\,u(D,v)/(2\Delta_u)\bigr)$ is $\eps$-DP. Some analyses use zero-concentrated DP
(zCDP) \citep{bun2016concentrated}, a relaxation based on R\'enyi divergence with a single
parameter $\rho$; costs add under composition, and $\rho$-zCDP implies
$\bigl(\rho+2\sqrt{\rho\ln(1/\delta)},\,\delta\bigr)$-DP for every $\delta\in(0,1)$.

Three properties determine how a guarantee behaves. Composition bounds repeated use: if
$\mathcal M_1,\ldots,\mathcal M_T$ are run adaptively on the same dataset and each is
$(\eps_t,\delta_t)$-DP for the same $\nb$ given the earlier outputs, then the joint transcript is
\citep{dwork2014algorithmic}
\begin{equation}
\label{eq:basic}
\Bigl(\,\sum_{t=1}^{T}\eps_t,\ \sum_{t=1}^{T}\delta_t\Bigr)\text{-DP}.
\end{equation}
Post-processing cannot weaken a guarantee: if $g$ depends on the private data only through
$\mathcal M(D)$, then $g(\mathcal M(D))$ has the same guarantee as $\mathcal M$. A computation that
later reads the raw data again is not post-processing. Group privacy extends a guarantee to inputs
connected by a chain of $m$ neighbouring steps. It gives $m\eps$ and
$\delta\sum_{j=0}^{m-1}e^{j\eps}$, an upper bound whose additive term grows geometrically.

A RAG system holds a corpus $D=\{z_1,\ldots,z_n\}$ of documents and a relevance score $r(z,q)$
\citep{lewis2020retrieval,karpukhin_dense_2020}. Given a query $q$, it retrieves the $k$
highest-scoring documents $R_k(q,D)$, and a frozen language model generates an answer $a$
conditioned on $q$ and $R_k(q,D)$. The corpus is private and the model is fixed independently of
it, so private data enter at inference time rather than through training, as in DP-SGD and
private fine-tuning \citep{abadi2016deep,li2022large}. Because an adversary
may choose each query
after seeing earlier answers, what must be protected is the whole transcript
$((q_1,a_1),\ldots,(q_T,a_T))$ under an adaptive query policy, together with anything else the
system exposes (Section~\ref{sec:release}).

\section{A Taxonomy of DP-RAG Approaches}
\label{sec:taxonomy}

Existing DP-RAG systems are best distinguished not by their mechanisms but by when they spend the
privacy budget: when each response is generated, continuously as queries arrive, or once before
any query is seen. This choice determines what the resulting $\eps$ protects, and we use it to
divide the literature into three families. Table~\ref{tab:families} records what each source paper
states and evaluates.

All five systems protect documents, but they do not use the same neighbouring relation
(Table~\ref{tab:families}, first row). MURAG and DP-SynRAG add or remove one document, and DP-RAG
one privacy unit, which it equates with a document. \citet{koga2024dprag} instead
replace one individual's data in a corpus of fixed size $n$, and DP-KSA's definition says only
``one data point'' while its proof uses replacement. We do not convert between these relations
implicitly. By group privacy, add/remove at $(\eps,\delta)$ implies replace-one at
$(2\eps,(1+e^{\eps})\delta)$, while the converse requires a mechanism-specific argument, for
example that removing a document is the same as replacing it with a null document that is never
retrieved. Below, $\nb_{\mathrm{doc}}$ denotes each paper's own relation, so comparisons across columns of Table~\ref{tab:families} carry the generic conversion $(2\varepsilon,(1+e^{\varepsilon})\delta)$. At DP-SynRAG's $(10,10^{-3})$ its $\delta$ is $22.0$, so the converted statement is vacuous.%
\begin{table}[t]
\centering\small
\setlength{\tabcolsep}{3pt}
\renewcommand{\arraystretch}{1.04}
\newcolumntype{Y}{>{\raggedright\arraybackslash}p{1.08in}}
\begin{tabular}{@{}>{\raggedright\arraybackslash}p{0.66in}YYYYY@{}}
\toprule
 & \multicolumn{3}{c}{\emph{query-time privatization}} & \emph{multi-query} & \emph{one-time corpus} \\
 \hline 
 & DPSparseVoteRAG \citep{koga2024dprag} & DP-RAG \citep{grislain2024ragdp}
 & DP-KSA \citep{tang2026dpksa} & MURAG(-ADA) \citep{wu2026murag} & DP-SynRAG \citep{mori2025dpsynrag} \\
\midrule
Relation
 & replace one individual's data, $|D|=n$ (\S II-B)
 & add/remove one privacy unit ($=$ one doc.)
 & ``differ in one data point'' (Def.~2.1)
 & add/remove one document (\S 2)
 & add/remove one record (\S 3.1) \\
\hline Unit
 & one person per doc.; multi-record case noted (fn.~1)
 & one unit per doc.; merge a unit's content (fn.~4)
 & \multicolumn{3}{c}{ownership of documents not discussed} \\
\hline Guarantee
 & $(\eps_q,\delta_q)$ per answer (Thm.~3)
 & $(\eps_q,\delta_q)$ per answer via an accountant
 & $(\eps_q,\delta_q)$ per answer (Thm.~5.1)
 & pure $M\eps_q$ for the stream (Thms.~1--2)
 & $(\eps_0,\delta_0)$ for $\tilde D$ (Thm.~1) \\
\hline Reported
 & $\eps_q\in[2,40]$, $\delta_q=10^{-4}$
 & $\eps_q\approx5$, $\delta_q=10^{-3}$
 & $\eps_q\in[1,8]$, $\delta_q=10^{-4}$
 & $\eps=10$, $M=1$
 & $\eps_0=10$, $\delta_0=10^{-3}$ \\
\hline Observable
 & answer & answer & keywords; answer is post-processing & answers & $\tilde D$ and all derived outputs \\
\hline Attack
 & S$^2$MIA & -- & -- 
 & Interrogation%
 & name leakage over 1100 queries \\
\bottomrule
\end{tabular}
\caption{DP-RAG systems as their source papers state and evaluate them (section, theorem, and
figure numbers refer to the source). $(\eps_q,\delta_q)$ is a per-response guarantee, $M$ bounds
the per-query charges to a document, $(\eps_0,\delta_0)$ is the guarantee of an offline release,
and online systems keep retrieval internal.
}
\label{tab:families}
\end{table}

\subsection{Query-time privatization}
\label{sec:family-querytime}

The first family leaves the corpus unchanged and privatizes each response as it is generated,
using the sample-and-aggregate framework \citep{nissim_smooth_2007}: the retrieved documents are
split into disjoint groups, the model is run once on each group, and only a noisy aggregate is
released.

In \citet{koga2024dprag}, the retrieved documents are partitioned uniformly at random into equal
groups that each vote for the next token; aligning the random permutations of neighbouring
retrievals changes at most one vote, which a rank-based grouping would not guarantee. Releasing the majority token privately at every position with the LimitedDomain
mechanism of \citet{durfee2019practical} makes the cost grow with answer length. DPSparseVoteRAG
reduces this dependence with the sparse vector technique \citep{dwork2014algorithmic}. A noisy test
checks whether the groups already support the token that the model would produce without
retrieval, and a private token is released only when they do not. \citet{koga2024dprag} compose
the per-token costs into an $(\eps_q,\delta_q)$ guarantee per answer. In the pure-DP variant
analysed by \citet{wu2026murag}, each private release and the tests before it cost
$\eps_{\mathrm{tok}}$, so $\eps_q\le c\,\eps_{\mathrm{tok}}$ for at most $c$ releases, however long
the answer.

\citet{grislain2024ragdp} privatizes both retrieval and generation. An exponential mechanism picks
a relevance threshold that roughly $k$ documents pass, and each token is sampled from an
exponential mechanism over clipped per-document log-probabilities combined with the model's public
next-token distribution. For an answer of $L$ tokens this gives
$\eps_q\le\eps_{\mathrm{ret}}+\sum_{j=1}^{L}\eps_{\mathrm{gen},j}$. DP-KSA \citep{tang2026dpksa}
privatizes keywords instead of the answer. It generates one response per
retrieved document, privately chooses a cut-off $\widehat k$ at a large gap between consecutive
token counts, and releases the $\widehat k$ most frequent tokens as keywords using
propose-test-release \citep{dwork2009robust}, falling back to a zero-shot answer if the stability
test fails. The final answer is generated from the keywords alone, which is post-processing.

This family spends budget only on decisions that require the sensitive documents, which gives
relatively good utility at a fixed per-answer budget: DPSparseVoteRAG outperforms a non-RAG
baseline mostly for $\eps_q\ge10$ \citep{koga2024dprag}, and DP-KSA from $\eps_q\approx2$
\citep{tang2026dpksa}. However, each guarantee concerns the map $D\mapsto a$ for one fixed query,
and over many adaptively chosen queries the per-answer guarantees compose.

\subsection{Multi-query accounting}
\label{sec:family-multiquery}

This family keeps the original corpus online and assigns the budget to records. Individual
accounting \citep{feldman2021individual} tracks a separate budget for each record, and a privacy
filter stops using a record before its accumulated charges exceed that budget
\citep{rogers2016privacy,feldman2021individual}. MURAG \citep{wu2026murag} gives each document a
budget $E_1(z)=M\eps_q$ and processes query $q_t$ as
\begin{equation}
\label{eq:murag}
\begin{aligned}
A_t &= \{z\in D : E_t(z)\ge \eps_q\}, &\qquad
\Dq{t} &= \{z\in A_t : r(z,q_t)>\tau\},\\
a_t &= \textsc{DP-RAG}_{\eps_q}\!\big(q_t,\ \topk_k(\Dq{t})\big), &\qquad
E_{t+1}(z) &= E_t(z)-\eps_q\,\ind{z\in \Dq{t}} ,
\end{aligned}
\end{equation}
where $\textsc{DP-RAG}_{\eps_q}$ is a pure-DP single-query mechanism, such as the variant above,
and $\tau$ is a fixed relevance threshold. Whether $z$ passes a fixed threshold depends only on its
own score and budget, so a document below it does not affect the $t$-th output. The filter of
\citet{feldman2021individual} needs exactly such a per-step bound on individual privacy loss given
the past transcript, here $\eps_q\,\ind{z\in\Dq{t}}$; at R\'enyi order $\infty$ it gives, for any
number $T$ of adaptively chosen queries,
\begin{equation}
\label{eq:murag-eps}
\sum_{t=1}^{T}\eps_q\,\ind{z\in\Dq{t}}\;\le\;M\eps_q
\quad\text{for every }z\in D,
\qquad\text{hence}\qquad
\eps_{\mathrm{stream}}\le M\eps_q .
\end{equation}
A document is charged whenever it passes the threshold, even if the top-$k$ step then drops it, so
a loose $\tau$ wastes budget. MURAG-ADA instead spends part of $\eps_q$ on a private,
query-specific threshold, scanning noisy counts over relevance bins until about $k$ documents have
been found, and charges every scanned document
$\eps_q=\eps_{\mathrm{thr}}+\eps_{\mathrm{RAG}}$, kept by top-$k$ or not. The underlying tools are filters and odometers
\citep{rogers2016privacy,feldman2021individual,lecuyer2021practical}. Individual accounting under
Gaussian DP and fully adaptive composition
\citep{koskela2022individual,smith2022fully,whitehouse2023fully} are related but are not used in
MURAG's stated pure-DP guarantee.

In MURAG, exhausting a budget retires a document instead of stopping the system. The active set
$A_t$ shrinks as its most-used documents are exhausted, at a rate set by the retrieval distribution
rather than by the number of queries. A document that is never charged costs nothing; one charged
by every query is exhausted after $M$ queries.

\subsection{One-time corpus privatization}
\label{sec:family-corpus}

The third family spends its budget once, offline. DP-SynRAG \citep{mori2025dpsynrag} releases a
DP synthetic corpus $\tilde D$ and answers queries using it. It builds on private
prediction \citep{amin2024private}: clipped per-document logits for a rephrasing prompt are summed
over a subset of documents, and the next token is sampled from their softmax, an exponential
mechanism. Because random subsets capture only coarse properties,
DP-SynRAG first clusters documents with a noisy keyword histogram, privately removes outliers as in
\citet{grislain2024ragdp}, and generates text from every cluster. A document may belong to up
to $\Lambda$ clusters, and in zCDP
\begin{equation}
\label{eq:synrag-eps}
\rho \;=\; \rho_{\mathrm{hist}}+\Lambda\big(\rho_{\mathrm{retr}}+\rho_{\mathrm{pred}}\big),
\qquad
\eps_0 \;=\; \rho+2\sqrt{\rho\ln(1/\delta_0)} ,
\end{equation}
where the terms are the costs of the histogram, the outlier removal, and the token sampling.
Related generators differ in how they form the aggregate
\citep{amin2025clustering,vinod2025invisibleink}, and heuristic variants use synthetic
data without a formal guarantee. Each answer
$a_t=\LLM{q_t,R_k(q_t,\tilde D)}$ is a function of an already private object, so by
post-processing the guarantee stays $(\eps_0,\delta_0)$ for every $T$, as long as serving never
accesses the raw corpus. In exchange, anything $\tilde D$ failed to preserve is unavailable to all
later queries.

\section{Three Dimensions for Interpreting Guarantees}
\label{sec:dimensions}

All three families report budgets of similar size. Written out in full, however, a DP-RAG
guarantee is the statement

\begingroup
\setlength{\belowdisplayskip}{4pt}

\begin{equation}
\label{eq:triple}
\forall\,\mathcal A\in\mathcal C,\ \ \forall\, D \nb D',\ \ \forall\, S:\qquad
\Pr\big[\mathcal{O}^{\mathcal A}_T(D)\in S\big] \;\le\;
e^{\eps(T)}\,\Pr\big[\mathcal{O}^{\mathcal A}_T(D')\in S\big] + \delta(T),
\end{equation}
\endgroup
where $\mathcal A$ is a query policy from the allowed class $\mathcal C$ (each query may depend on
earlier outputs), $\mathcal{O}^{\mathcal A}_T(D)$ is everything the adversary observes after $T$
queries chosen by $\mathcal A$, and probabilities are over the randomness of the mechanism and of
$\mathcal A$. The statement has three free slots: the relation $\nb$, the function
$T\mapsto(\eps(T),\delta(T))$, and the observable $\mathcal{O}$. A paper that reports
``$\eps=10$'' has given one value of one of them, and fixing one slot does not determine the other
two.

\subsection{Privacy granularity}
\label{sec:granularity}

Every system in Table~\ref{tab:families} is analysed at the level of documents, the retrieval
unit, which need not match the person a deployment protects. Let $D_u\subseteq D$ be the
documents of person $u$, and write $D\nb_{\mathrm{ind}}D'$ when one corpus is obtained from the
other by removing (for a replacement relation, replacing) all of some $D_u$. The two relations
coincide only when every person contributes one document. The source papers state the assumption
they need. \citet{koga2024dprag} ``assume that each RAG document comes from a single individual''
and note that individuals with several records can be handled by coarsening the relation, possibly
at a utility cost; \citet{grislain2024ragdp} suggests merging a privacy unit's content into
one document. What remains open is whether the assumption can be met and checked. Both
approaches need a map from documents to their owners, and merging changes the
retrieval unit itself: a patient's merged record is scored and retrieved as one document, which
changes what the generator sees.

Clinical corpora, the motivating application, need not have one record per patient. If
$|D_u|\le m$ for every person, group privacy turns a document-level guarantee into a person-level
one:
\begin{equation}
\label{eq:unit}
(\eps,\delta)\text{-DP w.r.t. }\nb_{\mathrm{doc}}
\quad\Longrightarrow\quad
\left(m\eps,\ \delta\sum_{j=0}^{m-1}e^{j\eps}\right)
\text{-DP w.r.t. }\nb_{\mathrm{ind}}.
\end{equation}
This generic upper bound holds for every mechanism; it is not the person-level privacy of any
particular one, which a mechanism-specific analysis may bound more tightly. For a
patient with twenty records under document-level $\eps=10$, it gives only
$\eps_{\mathrm{ind}}\le200$, and its $\delta$ becomes vacuous even sooner: at DPSparseVoteRAG's
$\eps=10$ and $\delta=10^{-4}$, two documents already give
$\delta_{\mathrm{ind}}=10^{-4}(1+e^{10})\approx2.2$. Pure DP converts to $m\eps$, so
the filter bound~\eqref{eq:murag-eps} becomes $mM\eps_q$, and $\rho$-zCDP becomes $m^2\rho$-zCDP
\citep{bun2016concentrated}.

The conversion needs a bound on $m$, and the benchmarks do not provide one. Encyclopedic corpora
have no contributors to count, and ChatDoctor \citep{li2023chatdoctor} is evaluated with each
patient--doctor conversation as one document \citep{koga2024dprag,wu2026murag}, without saying how
many belong to one person. Private training has faced the same gap
between example-level and user-level privacy \citep{charles2024userlevel}. Interpreting a reported
$\eps$ requires naming the protected unit and bounding per-person contributions
\citep{desfontaines2019sok}.

\subsection{Accounting scope}
\label{sec:scope}

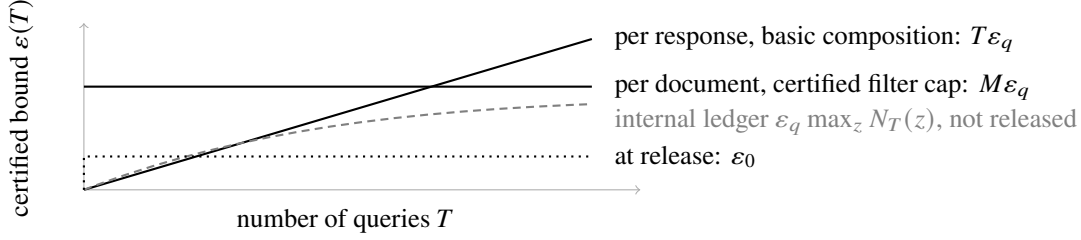
\begin{figure}[t]
\centering
\begin{tikzpicture}[x=1.28cm,y=0.42cm,font=\small]
  \draw[->,gray!65] (0,0) -- (5.75,0);
  \draw[->,gray!65] (0,0) -- (0,5.25);
  \node[font=\small,anchor=north] at (2.7,-0.28) {number of queries $T$};
  \node[font=\small,rotate=90,anchor=south] at (-0.42,2.6) {certified bound $\eps(T)$};
  \draw[thick] (0,0) -- (5.25,4.75);
  \node[anchor=west] at (5.38,4.75) {per response, basic composition: $T\eps_q$};
  \draw[thick] (0,3.25) -- (5.25,3.25);
  \node[anchor=west] at (5.38,3.25) {per document, certified filter cap: $M\eps_q$};
  \draw[thick,densely dashed,gray] (0,0) .. controls (1.5,1.75) and (3.1,2.45) .. (5.25,2.70);
  \node[anchor=west,gray] at (5.38,2.15) {internal ledger $\eps_q\max_z N_T(z)$, not released};
  \draw[thick,dotted] (0,0) -- (0,1.05) -- (5.25,1.05);
  \node[anchor=west] at (5.38,1.05) {at release: $\eps_0$};
\end{tikzpicture}
\caption{Three accounting scopes (Eqs.~\ref{eq:scope-resp}--\ref{eq:scope-rel}). Solid and dotted
lines are certified bounds. The dashed curve is the ledger that the filter keeps internally; it
depends on the corpus through retrieval. The curves are
schematic.}
\label{fig:scope}
\end{figure}

A budget reported for one response is not a deployment-wide budget. The second slot
of~\eqref{eq:triple} describes how the certified parameters grow with the number of queries, and
the families fill it differently (Figure~\ref{fig:scope}). For per-response guarantees, basic
composition~\eqref{eq:basic} gives
\begin{equation}
\label{eq:scope-resp}
\eps_{\mathrm{resp}}(T)\le T\eps_q,
\qquad
\delta_{\mathrm{resp}}(T)\le T\delta_q.
\end{equation}
Whether a better generic bound exists depends on $\eps_q$. Advanced composition
\citep{dwork_boosting_2010-1} does not help once $\eps_q\ge\ln2$, because its additive term
$T\eps_q(e^{\eps_q}-1)$ alone is at least $T\eps_q$. The bounds of
\citet{kairouz_composition_2017-1} are tighter. For $T=100$ pure $\eps_q$-DP responses and total
$\delta=10^{-6}$, their closed form (Thm.~3.4) gives $98.8$ instead of $100$ at $\eps_q=1$, and
their optimal composition (Thm.~3.3) about $83.5$. This optimum is tight over all
$\eps_q$-DP mechanisms, but it still gives $199.6$ of $200$ at $\eps_q=2$ and comes within
$10^{-3}$ of
$T\eps_q$ for $\eps_q\ge3$. The optimum also covers approximate DP: at $(10,10^{-4})$, basic
composition gives $(1000,\,0.01)$ for 100 responses and the optimum at the same $\delta$ is less
than $10^{-4}$ smaller. Optimal composition thus gains less than $10^{-3}$ at the
$\eps_q\approx5$--$10$ of Table~\ref{tab:families}, but gives about $192$ instead of $200$ at
DP-KSA's $(2,10^{-4})$.
Accounting directly for the underlying token-level mechanisms is mechanism-specific and
may give a smaller value, but no per-response paper states a stream-level bound of either kind.

Filters instead keep a budget for each document. Writing $\Dq{t}$ for the documents charged at
query $t$,
\begin{equation}
\label{eq:scope-doc}
N_T(z):=\sum_{t=1}^{T}\ind{z\in\Dq{t}}\le\min\{T,M\}
\quad\text{for every }z\in D,
\qquad
\eps_{\mathrm{doc}}(T)\le M\eps_q.
\end{equation}
The realised count $N_T(z)$ depends on the private corpus through retrieval, so it is an internal
ledger. The certified bound is the filter
guarantee~\eqref{eq:murag-eps} on the right, which needs a pure-DP per-query mechanism.
The cap $M$ replaces the factor $T$ and is enforced by retiring documents
(Section~\ref{sec:practice}). Finally, for an offline release that is $(\eps_0,\delta_0)$-DP and
is queried only through the artifact and public inputs, post-processing gives
\begin{equation}
\label{eq:scope-rel}
\eps_{\mathrm{rel}}(T)\le\eps_0,
\qquad
\delta_{\mathrm{rel}}(T)\le\delta_0
\qquad\text{for all }T.
\end{equation}

All three bound the document-level distinguishability of the $T$-query transcript, but for
different mechanisms. The first grows with the number of answers, the second is capped by $M$, and
the third stays constant. A per-response $\eps_q$ is the first curve at $T=1$ and can be compared
with the others only once $T$ is fixed; the flatter curves are paid for in utility. Amplification
by subsampling \citep{balle2018subsampling} requires a subsample drawn independently of the data,
as in private prompt-tuning and in-context learning methods
\citep{hong2024dpopt,duan2024flocks,tang2024privacypreserving,flemings2024differentially}.
Top-$k$ retrieval is a corpus-dependent ranking, so its cost cannot be discounted by
$k/|D|$. Poisson subsampling of the corpus at rate $\gamma$ does qualify and reduces the
per-response cost to $\log(1+\gamma(e^{\eps_q}-1))$, but the bound stays linear in $T$, and
the subsample is likely to miss the few documents each query needs; in \citet{wu2026murag}, this
construction stays below the non-RAG baseline.

\subsection{Release boundaries}
\label{sec:release}

The last dimension is the observable $\mathcal{O}_T$ in~\eqref{eq:triple}: the transcript for an
online system, the artifact $\tilde D$ for an offline one, and anything else the system exposes,
such as citations, retrieved passages, logs, or timing.

A transcript guarantee covers the answers only while retrieved documents and intermediate states
stay internal, as the threat models assume. Suppose an interface
also shows its sources, and let $S_z$ be the event that it releases an identifiable raw document
$z$. Under $D'=D\setminus\{z\}$ we have $\Pr[S_z\mid D']=0$, so~\eqref{eq:dp} requires
$\Pr[S_z\mid D]\le\delta$ for any value of $\eps$. Displaying $R_k(q,D)$ deterministically gives
$\Pr[S_z\mid D]=1$ for a retrieved $z$ and forces $\delta\ge1$. Randomising which genuine documents
are shown does not fix this, because DP still allows each identifiable document to be shown
verbatim with probability at most $\delta$. Showing sources therefore requires sanitised
citations, an output space without identifiable raw text, or a different disclosure model, each
with its own analysis.

For an offline system, any index, retrieved passages, and answers computed only from $\tilde D$
are post-processing, until some computation reads $D$ again. Protecting the artifact thus protects
every transcript computed only from it, under the artifact's adjacency relation and parameters.
This does not rank arbitrary offline systems above or below online ones, and the price is that
fidelity is fixed before any query arrives. Protecting the user's query instead of the corpus is
another boundary, with a different adversary \citep{cheng2024remoterag}. A reported $\eps$ should
therefore state which observable it covers.

\section{What Complicates Comparison in Practice}
\label{sec:practice}

Fixing the three slots of~\eqref{eq:triple} makes two guarantees comparable in principle, but not
two reported evaluations, because most quantities that determine utility under a privacy constraint
go unreported.

The first is the query distribution. When budget is charged to documents rather than queries, how
often the same documents are needed determines how quickly access to the corpus is lost. A
deliberately simple model makes this concrete. Suppose queries are drawn independently from a
distribution $Q$, and let $p_z=\Pr_{q\sim Q}[r(z,q)>\tau]$ be the probability that a query charges
document $z$ (for MURAG-ADA, that the private scan reaches $z$'s bin). Under the
filter~\eqref{eq:murag}, $z$ is still active at query $t$ exactly when it
has been charged fewer than $M$ times, so
\begin{equation}
\label{eq:survival}
\pi_t(z) \;:=\; \Pr[z\in A_t] \;=\;
\sum_{j=0}^{\min\{M-1,\,t-1\}}\binom{t-1}{j}\,p_z^{\,j}\,(1-p_z)^{t-1-j},
\qquad\text{and for } M=1:\quad \pi_t(z)=(1-p_z)^{t-1} .
\end{equation}
At $M=1$, the setting of the main experiments in \citet{wu2026murag}, a document has a half-life of about seventy queries for $p_z=0.01$ and
fewer than three for $p_z=0.25$. The same $\eps=M\eps_q$ can therefore describe very different
systems depending on $\{p_z\}$. Independent queries are not the favourable case in general: a
stream deliberately spread over disjoint sets of documents would retire nothing a later query
needs. Access is lost through repeated overlap on the same documents, as in correlated
streams whose $t$-th query needs documents that earlier queries have spent. MURAG-ADA was designed
around this effect, and among the evaluations in Table~\ref{tab:families} only \citet{wu2026murag}
measure document reuse: most documents are retrieved for one or two of a hundred open-domain
questions \citep{kwiatkowski_natural_2019,joshi_triviaqa_2017},
whereas many recur across multi-hop chains. Non-overlapping benchmark
questions favour this family, while related questions about the same records, closer to the
clinical use case, put it under pressure, so the overlap of the query set should be reported with
the results.

Where in the stream utility is measured can also complicate comparison. Because retirement can remove
documents that a later query needs, answer quality can become a function $U(t)$ of
position in the stream. Eq.~\eqref{eq:survival} describes when documents are retired but does not
by itself imply that quality falls monotonically with survival, since a query may still be answered
from surviving documents or from the model's own knowledge. The reported mean
$\bar U_T=\frac{1}{T}\sum_{t=1}^{T}U(t)$ averages answers given before and after the most relevant
documents are retired, so equal means can hide different curves. The other two families never
retire documents, so under stationary evaluation conditions they avoid this history-dependent loss
of access, although their $U(t)$ can still vary with query difficulty, distribution shift,
adaptive queries, or conversation history. Plotting quality against query index separates the two
effects and costs nothing extra.

The third is whether attack evaluations match the guarantees. Attacks on unprotected RAG report
empirical success for a specific adversary, corpus, retriever, and generator
\citep{anderson2024ismydata,liu2024maskmia,naseh2025riddle,li2025generating,jiang2024feedback,peng_data_2024}.
Two of the defences also evaluate membership inference (Table~\ref{tab:families}):
\citet{koga2024dprag} apply S$^2$MIA \citep{li2025generating} with one query per target, and
\citet{wu2026murag} the Interrogation Attack \citep{naseh2025riddle} with thirty. At
$\eps=10$, both report AUCs of $0.45$--$0.53$, compared with $0.61$--$0.66$ without protection. The
relevant bound follows from~\eqref{eq:ht}: for a fixed target $z$ and any certified pair
$D\nb_{\mathrm{doc}}D'$ with $z\in D\setminus D'$, any test between them satisfies
\begin{equation}
\label{eq:adv-bound}
\mathrm{TPR} \;\le\; \min\!\big\{\,e^{\eps}\,\mathrm{FPR}+\delta,\ \
1-e^{-\eps}(1-\delta-\mathrm{FPR})\big\},
\qquad
\mathrm{TPR}-\mathrm{FPR} \;\le\; \frac{e^{\eps}-1+2\delta}{e^{\eps}+1} .
\end{equation}
Under add/remove (DP-RAG, MURAG, DP-SynRAG) $D'=D\setminus\{z\}$; under replace-one
(DPSparseVoteRAG, DP-KSA) $D'$ swaps $z$ for some $z'$, so a deletion game lies outside these
guarantees without a mechanism-specific argument. With $\delta=0$, the advantage (a single point on the ROC curve) is at most $0.46$ at
$\eps=1$ and $0.9999$ at $\eps=10$. Integrating the TPR bound gives
$\mathrm{AUC}\le e^{\eps}/(1+e^{\eps})$, that is, $0.73$ and $0.99995$. An AUC near $0.5$ is thus
well within what $\eps=10$ allows: it is evidence about the mechanism on that corpus, not a
consequence of the guarantee. Two conditions determine whether the comparison is
meaningful. First, the bound~\eqref{eq:adv-bound} concerns neighbouring corpora and a fixed
target, whereas both evaluations pool members $D_{\mathrm{in}}\subset D$ with a disjoint set
$D_{\mathrm{out}}$ that both papers describe as out-of-distribution. Unless the two populations are
exchangeable, an attack can separate them by content alone, and the pooled AUC then reflects this
mismatch as well as membership. Second, the $\eps$ must be the one in force for the attacker's
queries. A thirty-query attack on a per-response system is covered only by the basic-composition
bound $\eps_{\mathrm{resp}}(30)\le30\,\eps_q$, whereas under MURAG the reported $\eps=M\eps_q$
already covers all thirty queries. Papers usually leave implicit how each attack's adjacency, query
budget, and observable match the guarantee.

A common convention underlies all three issues: evaluating near $\eps=10$, the main operating
point of three of the five systems in Table~\ref{tab:families}. We found no deployment-based
justification for this value in these papers: \citet{koga2024dprag} report gains over the non-RAG
baseline mostly for $\eps_q\ge10$, and \citet{wu2026murag} follow their setup. This value allows a
likelihood ratio of $e^{10}\approx2.2\times10^{4}$ in~\eqref{eq:dp}; it reflects where the methods
become useful.%

\section{Conclusion}
\label{sec:open}
\label{sec:conclusion}

Privacy accounting in RAG has moved from protecting a single answer to protecting a running
service, but budget reporting has not kept up. A reported $\eps$ is
meaningful only together with the three slots of~\eqref{eq:triple}: the protected unit, the
accounting scope, and the release boundary. The same nominal $\eps=10$ means a per-response
$\eps_q$ with basic-composition bound $T\eps_q$ in the first family, a per-document cap $M\eps_q$
in the second, and a one-time $\eps_0$ in the third. For a person with $m$ documents, the generic
conversion multiplies the certified $\eps$ by $m$. Before guarantees in this literature can be
compared, papers should report these quantities together with the evidence in
Table~\ref{tab:families}, from the original neighbouring relation and theorem to query overlap and
the number of queries per attacked target.

{\small
\bibliographystyle{abbrvnat}
\bibliography{main}
}

\end{document}